\documentclass[printer]{aa}

\usepackage{graphicx}
\usepackage{txfonts}
\newcommand{\logm}{log M$_\star$/M$_\odot$ $\sim$ }
\begin{document} 

   \title{Sub-Gyr variability around the SFMS and its contribution to the scatter}
   \titlerunning{Sub-Gyr variability around the SFMS and its contribution to the scatter}

   \author{A. Camps-Fariña
          \inst{1,2}\fnmsep\thanks{\email{arcamps@ucm.es}},
          M. Chamorro-Cazorla\inst{1,2}
          \and
          S. F. S\'{a}nchez\inst{3,4}
          }
    \authorrunning{A. Camps-Fariña et al.}

   \institute{Departamento de F\'{i}sica de la Tierra y Astrof\'{i}sica, Universidad Complutense de Madrid, Pl. Ciencias, 1, Madrid, 28040, Madrid, Spain
   \and
   Instituto de Física de Partículas y del Cosmos, Universidad Complutense de Madrid, Pl. Ciencias, 1, Madrid, 28040, Madrid, Spain
   \and
   Instituto de Astronom\'{i}a, Universidad Nacional Aut\'{o}noma de M\'{e}xico, A.P. 106, Ensenada 22800, BC, M\'{e}xico
   \and
   Instituto de Astrof\'{i}sica de Canarias, V\'{i}a L\'{a}ctea s/n, 38205 La Laguna, Tenerife, Spain
             }
   \date{Received September 15, 1996; accepted March 16, 1997}

% \abstract{}{}{}{}{} 
% 5 {} token are mandatory
 
  \abstract
   {}
   {We aim to measure the evolution of individual galaxies around the Star Formation Main Sequence (SFMS) during the last Gyr as a function of their stellar mass to quantify how much of its scatter is due to short-term variability.}
   {We derived star formation histories using full spectral fitting for a sample of 8,960 galaxies from the MaNGA survey to track the position of the galaxies in the SFMS during the last Gyr.}
   {The variability correlates with both the stellar mass of the galaxies and their current position in both the SFMS and the mass-metallicity relation (MZR), with the position in the latter strongly affecting variability in SFR. While most of the fluctuations are compatible with stochasticity, there is a very weak but statistically significant preference for $\sim$$135-150$ Myr time-scales.}
   {These results support a strong self-regulation of SFR within galaxies, establishing characteristic intensities and time-scales for bursts of star formation and quenching episodes. We also find that short-term variability cannot account for the entirety of the scatter in the SFMS. It appears to originate to a similar degree in short-term variability and long-term (halo-level) differentiation and fits predictions from models.}
  {}% conclusions heading (optional), leave it empty if necessary 

   \maketitle
%

%-------------------------------------------------------------------
\section{Introduction}
The Star Formation Main Sequence (SFMS) and the Mass-Metallicity Relation (MZR) are two of the most widely studied relations in galaxy evolution. Their very existence implies a link between the current properties of the galaxies and their cumulative history as both, the stellar mass (M$_\star$) and metallicity are the direct consequence of the past star formation history (SFH).

Many authors have interpreted the low scatter among the SFMS ($\sim$0.3 dex), observed at different redshifts, as evidence that star formation in disk galaxies is regulated by feedback processes and gas accretion \citep[e.g.,][]{Whitaker2012,Ilbert2015,Salmon2015,Popesso2019}. Alternatively, \cite{Kelson2014} argues that the relation and its scatter can be reproduced using stochastic prescriptions for the SFH as a random walk of the star formation rate (SFR), and that it arises from the central limit theorem.

Several works have studied secondary correlations of the SFMS to assess which mechanisms preferentially regulate the growth of galaxies. \cite{GonzalezDelgado2016,Cano-Diaz2019} propose morphology as partly responsible for the scatter of the SFMS both in its global and resolved versions. \cite{Saintonge2016} find similar results regarding gas fraction (which strongly correlates with morphology) but \cite{Hall2018} find no effect of morphology or gas fraction in the resolved SFMS but instead a critical mass above and below which different mechanisms contribute to the scatter. \cite{Barrera-Ballesteros2021, Ellison2024} favor instead gas pressure as a factor contributing to the scatter and therefore, regulating the efficiency of star-formation. On global scales, \cite{Berti2021} ties the scatter to the large-scale environment and \cite{Gladders2013,Rodriguez-Puebla2016,Matthee2019,Blank2021} propose that most of the scatter is due to differences in formation time of the parent halos. This means that it is not intrinsically a difference in SFR values that creates the scatter. Instead, some galaxies have lower or higher M$\star$ values due to having different life-times to accumulate M$_\star$ for a given halo mass (which drives gas accretion rates and therefore SFR).

These mechanisms are expected to act upon a variety of temporal scales, meaning that one of the ways to discriminate between them is to check whether the scatter results from short-term SFR variations or if the current position of a galaxy relative to the SFMS is maintained over longer time-scales in its lifetime. If galaxies are dominated by short-scale fluctuations ($<$1 Gyr) it means that the SFMS is intrinsically tighter than is observed and the scatter originates in short stochastic deviations such as random bursts or decreases in star-formation \citep[e.g.,][]{Peng2010,Behroozi2013,Speagle2014}.

Longer time-scale fluctuations of the SFMS (several Gyr) are typically associated to larger physical scales in the Universe, either the aforementioned delays in halo formation time in galaxies, or the environment via differences in the clustering of galaxies \citep{Berti2021}. In practice, these produce offsets or differences in shape between SFHs such that some galaxies spend most of their life-time in the main sequence above the SFMS and others preferentially stay below it. Alternatively, \cite{Morselli2017} show that variations perpendicular to the SFMS correlate with the morphology of the galaxies.

Measuring the motions of galaxies on the parameter space of the SFMS at different time-scales is very difficult. The variability of the SFMS has been mainly studied in cosmological simulations where the SFH of the galaxies can be readily obtained without worsening reliability at older ages as it happens with SED and full spectrum fitting techniques.
\cite{Tacchella2020, Iyer2020, Wang2020b} among others have studied the variability of the SFHs on different time-scales to find similar results to those of \cite{Matthee2019} showing that the temporal variation of the SFR in galaxies is a combination of longer time-scales with changes on the Gyr scale and shorter fluctuations of period 10-100 Myr consistent with the ages of star-forming regions and star-bursts.
Observationally, SFR variability or "burstiness" has only been studied using indirect methods, such as by comparing the SFR values measured using different tracers such as H$\alpha$ and UV fluxes which are associated with different time-scales and therefore the difference between their values can be considered to arise from short-term variability \citep[e.g.,][]{Weisz2012, Guo2016, Emami2019, Caplar2019, Wang2020a,Patel2023, Clarke2024}.

In this article we approach this topic from a purely observational viewpoint by measuring the number of times each galaxy crosses the SFMS in its recent history. We use SFHs obtained by full spectral fitting and focus on the better-sampled and more reliable last Gyr in age to determine how many times they have crossed the SFMS as well as how much they deviate from it after crossing them and how much time passes before they cross it again. The crossings are associated with short time variability (10-100 Myr) in the galaxies, and we use their properties and those of the host galaxy to discern which physical processes they could be associated with. We also assess whether these oscillations can account for the scatter in the relation or if galaxies maintain their position relative to the SFMS over longer time-scales.
In Sec. \ref{sec:data} we present the data employed and in \ref{sec:analysis} how it was processed to obtain the number of crossings. In Sec. \ref{sec:results} we present the results which we discuss in Sec. \ref{sec:discussion}.
We adopt a standard $\mathrm{\Lambda}$CDM cosmology (H$_\mathrm{O}=73$ km s$^{-1}$, $\mathrm{\Omega_M} = 0.3$, and $\mathrm{\Omega_\Lambda} = 0.7$) and a Salpeter IMF \citep{Salpeter1955} throughout this work.

\section{Data}\label{sec:data}
The MaNGA survey, as described by \cite[][]{Bundy2015,Abdurro'uf2022}, comprises integral field unit (IFU) spectroscopic observations conducted on a group of roughly 10,000 local galaxies selected based on their luminosity in the redshift range 0.01 < z < 0.15 ($\langle \mathrm{z} \rangle \sim 0.03$).
Data were acquired using the BOSS spectrographs \citep[][]{Smee2013} on the Sloan 2.5\,m telescope at Apache Point Observatory \citep{Gunn2006}. In addition to the fiber bundles used for the observation of each galaxy, varying in size depending on the observed object \citep{Drory2015}, additional fibers and fiber bundles were used for the flux calibration and sky subtraction tasks \citep{Yan2016}.
The data cubes, spanning a spectral range from 3,600\,$\AA$ to 10,300\,$\AA$ with a spectral resolution of approximately R$\sim$2,000 and a spatial resolution of roughly 2.5\,\arcsec/FWHM, were derived through the reduction and calibration of the observations using the Data Reduction Pipeline \citep[DRP,][]{Law2016}.

The full sample (DR17, 10,245) was refined following \cite{Camps-Farina2022} by selecting galaxies which passed the quality control described in Sec. 4.5 of \cite{Sanchez2022}. Our sample selection and subsequent analysis makes use of the pyPipe3D data produces which are described in detail in Sec. \ref{sec:pipe3d}. As a brief summary, first an automatic check is done to make sure that all data-products have been created and all the relevant scripts have been successfully executed, followed by comparing the values for the redshift and M$_\star$ that we recover with those of the NSA catalog\footnote{\url{http://nsatlas.org/}}. These are expected to be stable values regardless of the method employed to measure them, so galaxies exceeding a 30\% discrepancy are discarded. This is followed by a multiple-round human inspection of the central spectrum and its fit, mass assembly and chemical enrichment histories, mock photometry vs SDSS images, and maps of emission lines, ages, stellar abundances and diagnostics such as the BPT \citep{Baldwin1981}.
Since we measure the metallicity at the effective radius R$_\mathrm{e}$ (see next section), we additionally select only galaxies below 70º in inclination and remove optically detected AGN \citep[types I and II, see][]{Lacerda2022} to avoid contamination of their emission in the stellar spectra.
The final sample includes 8,960 galaxies spanning a range of stellar masses between $\log M_\star/M_\odot \sim 8.5 - 12$ as well as all morphological types.

\section{Analysis}\label{sec:analysis}
We use full spectral fitting to derive the SFH of the galaxies and trace their position in the SFMS during the last Gyr. Full spectral fitting is used to determine what fraction of light emitted today can be assigned to a set of stellar populations of given ages and metallicities. These fractions of light can be converted into the amount of M$_\star$ that was formed at a given age and the average abundance of the ISM at that time.

\subsection{pyPipe3D} \label{sec:pipe3d}
We derive SFHs using \texttt{pyPipe3D} \citep[][]{Sanchez2006,Sanchez2016a,Sanchez2016b,Lacerda2022} together with the MaStar-sLOG stellar population template library \citep{Sanchez2022}. \texttt{pyPipe3D} is a full spectrum fitting software capable of fitting both the absorption and emission components of galaxy spectra and analyzing them separately. The individual spaxels of the MaNGA data-cubes are first spatially binned to ensure a good S/N $\gtrsim$ 50. Each stellar spectrum is then analyzed using a non-parametric fitting procedure to find the linear combination of spectral models that best fits the observed spectrum. What this means is that no shape of the SFH is assumed, the pipeline only returns the light fractions for each individual population which are used to reconstruct the SFH. Therefore, any variability in the histories can only arise due to the intrinsic physical changes in the values or be due to noise and uncertainties in the measurements or fitting.

MaStar-sLOG is an SSP library constructed using spectra from the MaStar \citep{Yan2019} stellar library and Charlot-Bruzual 2019 models\footnote{\url{https://www.bruzual.org/CB19/}} made with the GALAXEV software \citep{Bruzual2011}. MaStar spectra were observed using the BOSS spectrographs which are the same as those employed in MaNGA, ensuring that the resolution and spectral coverage of the fitted and template spectra are identical. This setup is ideal for the purposes of stellar population fitting techniques and will reduce the systematic errors in the results. The library is comprised of 39 ages between 1 Myr and 13.5 Gyr using a pseudo-logarithmic spacing and 7 metallicity values (Z = 0.0001, 0.0005, 0.002, 0.008, 0.017, 0.03, and 0.04) \citep[see Sec. 4.2 in][]{Sanchez2022}.

pyPipe3D performs the fitting in several steps meant to improve the precision of the recovered results. First, the non-linear parameters (redshift, dust attenuation and velocity dispersion of the stellar component) are determined using a reduced set of SSP and spectral range to avoid degeneracies (see \cite{Sanchez-Blazquez2011}), with the emission lines masked. We employ a \cite{Cardelli1989} extinction law for the dust attenuation. Once these parameters are fixed, the resulting stellar continuum is subtracted and the emission lines fitted. The derived emission lines are then subtracted from the spectrum, leaving only the stellar component, which is then fitted with the full SSP templates. The fitting procedure includes a random exploration of the parameter space for the line-emission fitting as well as Monte-Carlo iterations with perturbations for the derivation of the SSP light fractions, ensuring the results are not affected by local minima in the parameter space and that they are as robust and reliable as possible. For more details, we refer the reader to \cite{Lacerda2022} for a description of the code and \cite{Sanchez2022}, where its application to MaNGA and its results are shown and discussed in detail. Our SFHs are derived from the light fractions obtained in \cite{Sanchez2022}. 

\begin{figure}%
\centering
\includegraphics[width=\linewidth]{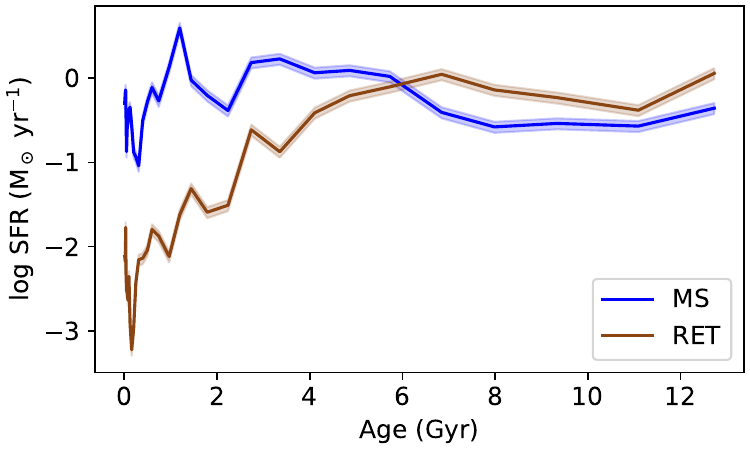}
\caption{Example SFHs obtained from the full spectral fitting performed on the MaNGA sample. We show one galaxy within the main sequence of the SFMS (MS) and one located over 1$\sigma$ below it (RET). The corresponding PLATEIFU identifiers of the data cubes are 12514-12701 for the MS galaxy and 10001-1902 for the RET galaxy.}\label{fig:hist_example}
\end{figure}

For the analysis presented in this work, we derive global values of the SFH for each galaxy, obtained by adding the SFHs of all the bins as well as the metallicity [Z/H] at the effective radius (R$_\mathrm{e}$) to derive the MZR. The [Z/H] at the R$_\mathrm{e}$ is calculated by fitting the radial gradient of metallicity at each age and selecting the value at the R$_e$. The metallicity at the R$_e$ has been shown to be a robust estimation of a galaxy's global metallicity \citep{GonzalezDelgado2014, Sanchez2020}.

Although the spatial coverage of the data is not exactly the same for all the galaxies, the MaNGA sample selection guarantees that we reach out to 1.5\,R$_e$ and 2.5\,R$_e$ for 80\% of the Primary and Secondary sub-samples, respectively, which does not significantly change our measurement of [Z/H] at the R$_\mathrm{e}$.
For the SFR and M$_\star$, however, this will impact their total value which can potentially affect the results in two ways, either in terms of the fluctuations of the SFH or inducing an artificial offset to the SFMS. The fluctuations in the SFH should not be affected significantly unless the bulk of star formation occurs at the outskirts of the galaxies which should be rare. The loss of light will decrease both the SFR and M$_\star$, meaning that the offset to the SFMS will be canceled out to a large extent. In appendix B of \cite{Camps-Farina2024} it is shown that the lost flux is relatively small (0-20\%) and fairly independent of M$_\star$ and in Appendix \ref{sec:app-primsec} of this article we show that the Primary and Secondary sub-samples do not show a significant offset in the SFMS and MZR.

In Fig. \ref{fig:hist_example} we show the SFH for two galaxies as an example, one of them located within 1$\sigma$ of the SFMS and one below this value. The galaxy currently below the SFMS used to be star-forming but eventually retired with a steep drop in SFR.

\begin{figure}
\centering
\includegraphics[width=\linewidth]{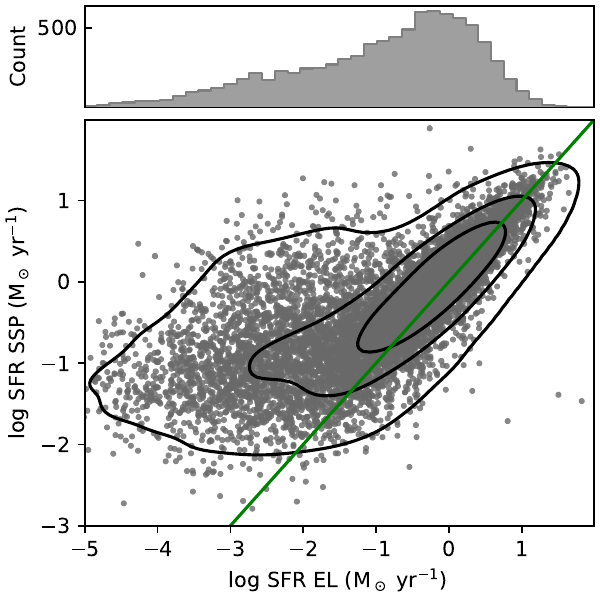}
\caption{Comparison between the SFR measured using emission lines (X-axis) and the full spectral fitting we employ selecting the last 30 Myr (Y-Axis) for all 8960 galaxies in our sample. The contours are set to encompass 95\%, 65\% and 35\% of the sample. The top panel shows the distribution of data-points.}\label{fig:app-match}
\end{figure}

The emission line spectra are not used to measure either the SFR or [Z/H] at any point in the article and are only used for comparison purposes. We do, however, use the EW$_\mathrm{H\alpha}$ to select star-forming galaxies and define the star-formation sequence in the SFMS \citep{Cano-Diaz2016}. We also use the emission lines for a consistency check on the reliability of the SFR values by comparing the current values derived from emission lines to equivalent measurements using the SFH in Fig. \ref{fig:app-match}. In making this comparison it is very important to ensure that the time-scales associated to each estimate of the abundance and the SFR match properly which for the case between full spectral fitting and optical emission lines is about 30 Myr \citep{Asari2007,GonzalezDelgado2016}. Therefore, the full spectral fitting values we show in the figure are measured by selecting only the light fractions for templates with ages under 30 Myr. The emission line SFR values are calculated from the dust-corrected H$_\alpha$ emission using the H$\alpha$/H$\beta$ ratio as described in \cite{Sanchez2022}.
The SFR values are taken as the global value for the galaxy and they follow the 1:1 relation quite well ($\sigma \sim 0.35$ dex at $\mathrm{log \; SFR \; EL}>-2$) except at very low values, which is also the case in which both the light fractions and the emission line fluxes will be very low and closer to the noise.

\subsection{Crossing the SFMS} \label{sec:cross_det}

\begin{figure*}%
\centering
\includegraphics[width=\linewidth]{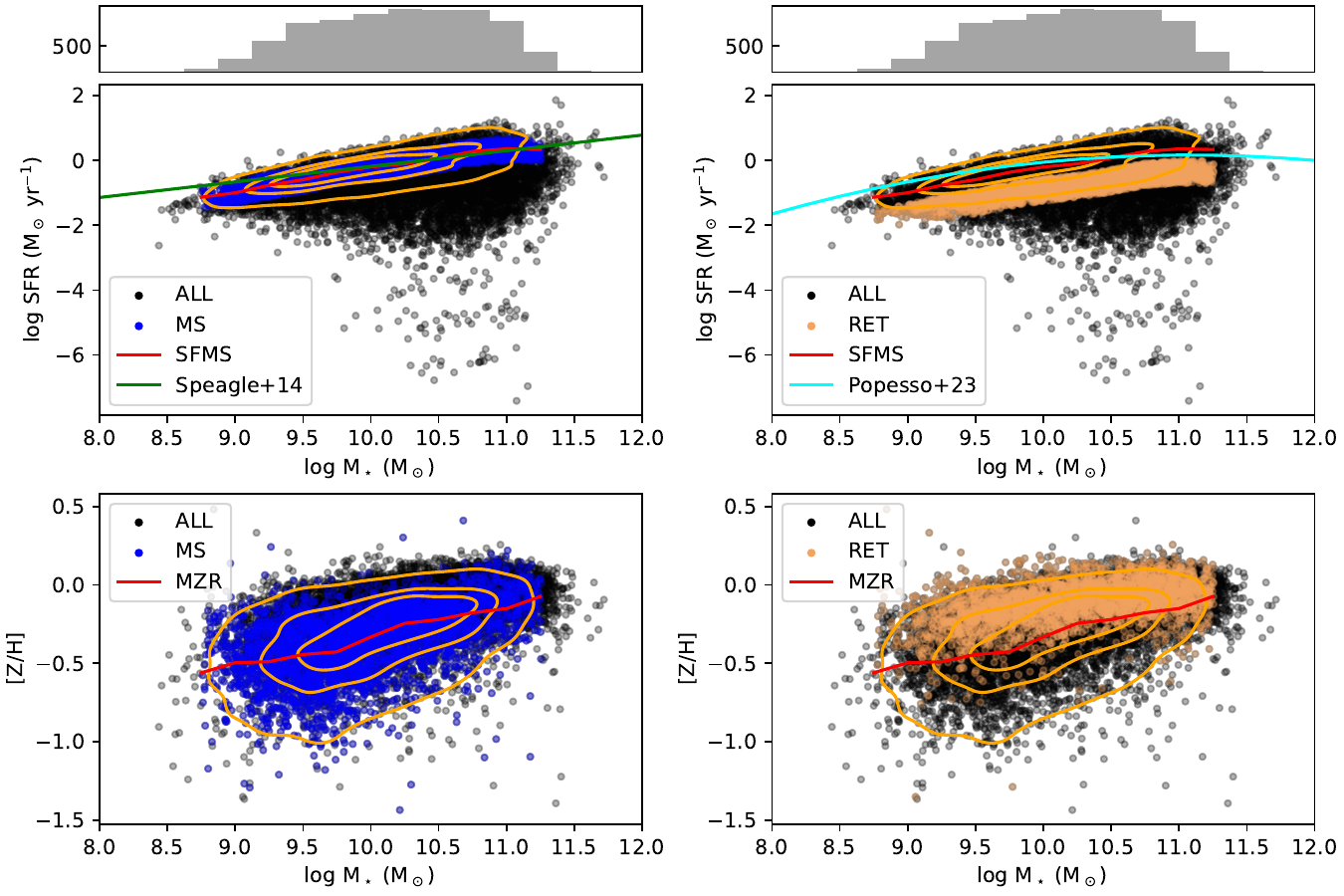}
\caption{SFMS and MZR measured from the most recent values of the SFHs and the [Z/H] we employ ($\sim$20 Myr) obtained by full spectral fitting. The full distribution (ALL) is shown as black points, with MS (blue) and RET (brown) shown in colored points in the left and right columns, respectively. The contours follow the star-forming sub-sample as defined in Sec. \ref{sec:cross_det} which is used to measure the SFMS and MZR (red lines). In the top two panels we show the SFMS at z$\sim0$ from \cite{Speagle2014} (left) and \cite{Popesso2023} (right) for comparison with ours.}\label{fig:current}
\end{figure*}

\begin{table}
\caption{Age bins within the last 1 Gyr. The bins within the first 20 Myr for the SFR and [Z/H] are averaged and assigned to their median age.}             % title of Table
\label{tab:agebins}      % is used to refer this table in the text
\centering                          % used for centering table
\begin{tabular}{r c c c}        % centered columns (4 columns)
\hline\hline                 % inserts double horizontal lines
& Age (Myr) & Age (Myr) & Age (Myr)\\    % table heading 
\hline                        % inserts single horizontal line
\multicolumn{1}{r|}{}& 1.5	&	29.3	&	201.2	\\
\multicolumn{1}{r|}{}& 3.0	&	37.4	&	248.7	\\
\multicolumn{1}{r|}{}& 4.7	&	47.7	&	310.2	\\
\multicolumn{1}{r|}{6.8 Myr}& 6.8	&	61.2	&	396.9	\\
\multicolumn{1}{r|}{}& 9.6	&	79.4	&	497.5	\\
\multicolumn{1}{r|}{}& 13.1	&	99.5	&	597.9	\\
\multicolumn{1}{r|}{}& 17.3	&	124.1	&	743.3	\\ 
& 22.8	&	158.7	&	967.0	\\

\hline                                   %inserts single line
\end{tabular}
\end{table}

To quantify the evolution of galaxies in the SFMS we calculate the number of times they cross it during the last Gyr, transitioning from having an SFR higher than predicted for its mass to values below, or vice versa.
The first step is to determine the SFMS as well as the MZR, which will be used to see if the crossings of the SFMS correlate with where galaxies are in the MZR. We are mostly interested in currently star-forming galaxies since we probe the recent history and therefore, we employed a cut selecting galaxies with EW$_\mathrm{H\alpha}>6\AA{}$ \citep{Cano-Diaz2016} to define the star-forming sample which we use to measure the SFMS and MZR. Since the SFMS evolves over time and we are measuring the position of galaxies relative to it we recalculate it at each age for the comparison. As such, we ensure that each crossing represents a change in the position of the galaxy with respect to the average value for currently star-forming galaxies. As seen in Fig. \ref{fig:current}, the SFMS for the most recent bin agrees well with the measurements from \cite{Speagle2014} and \cite{Popesso2023} at $\mathrm{z}\sim0$.

Due to the redshift range in MaNGA, the age values at which we measure the SFR correspond to different look-back times (LBT) when corrected for light-travel time. Ideally, we would apply this correction so that the histories are on the same cosmic clock, but this would introduce a selection bias in the sample as the LBT bins would not be equally populated. In fact, a significant number of galaxies are located far enough that their most recent measurement is older than 1 Gyr in LBT. Because of these inconsistencies it is better to use ages so that the results are comparable between galaxies at the expense of measuring galaxies at different cosmic times for the same age bin. Due to the selection strategy of MaNGA (which observes more luminous galaxies at higher z) this will mainly correlate with M$_\star$ in our results, meaning that massive galaxies are measured up to $\gtrsim$1 Gyr earlier. We do not expect this to cause a significant difference in the results, as galaxies with similar mass will have a very similar redshift. As a result, the SFMS and MZR will be largely consistent within M$_\star$ bins which is good enough for us to measure the crossings for each galaxy individually.

The number of crossings is determined as the number of times the sign of the difference between a galaxy's SFR and that of the SFMS changes within the past Gyr of each galaxy's history. In other words, each time that a galaxy goes from being above the relation to being below or vice-versa. An example of this is shown in Fig. \ref{fig:example}.
We select the last Gyr due to the sampling of the SSP library we employ, which has 24 values within the last Gyr but only 3 between 1 and 2 Gyr. This is because the spectra of SSPs change much more slowly with time at ages greater than 1 Gyr. The first few bins have very low age intervals ($\sim10-10$ Myr) on which galaxies are not expected to significantly vary their SFR, which is why we averaged the first 20 Myr of the SFH (see Table \ref{tab:agebins}).

We select three bins for the galaxies depending on their position in the SFMS: (i) all galaxies (ALL), (ii) galaxies within 1 $\mathrm{\sigma_{SFMS}}$ of the measured SFMS which are considered to be main sequence (MS) galaxies and (iii) a band of the same width (1 $\mathrm{\sigma_{SFMS}}$) as MS but immediately below it in the SFMS, defined as RET. The RET bin is meant to represent galaxies located below the SFMS. We imposed the same width as MS for similarity between the bins and to avoid galaxies which will have very low fractions of light in young populations, impacting the quality of the SFH and [Z/H] measured in the last 1 Gyr. There are 8960 galaxies in ALL, 4654 in MS and 2650 in RET.

In Fig. \ref{fig:current} we show the measured SFMS and MZR for the galaxies and how MS and RET are distributed in the relations.
In the SFMS, MS galaxies are located close to the relation for star-forming galaxies and RET are located immediately below them, as would be expected since these are the definitions employed to select them. In the MZR the MS galaxies are distributed over the entire parameter space that ALL occupy, though centered around the measured relation for star-forming galaxies. RET galaxies, on the other hand, are located in a tighter sequence above the MZR, showing that quenched galaxies have higher metallicities. However, the age-metallicity degeneracy can make it hard to distinguish between whether a galaxy has old ages or high abundances in their stellar populations, in Fig. \ref{fig:app-match} the galaxies with very low SFR (which will populate mostly the RET sample) appear to be offset to higher SFR as measured from their stellar populations compared to using emission lines. It is possible then that the higher metallicity for RET galaxies comes from these galaxies being measured with younger stellar populations than they should be. Since the measurements use very young stellar populations ($\lesssim20$ Myr) it is unlikely for the entire offset to be due to degeneracy since the correlation between age and metallicity largely breaks down for younger ages. This is at the expense of the latter being harder to measure due to its associated absorption lines being less visible in spectra dominated by very young stars \citep[e.g.,][]{Vazdekis2008,Conroy2013}.

\section{Results} \label{sec:results}
\begin{figure}%
\centering
\includegraphics[width=\linewidth]{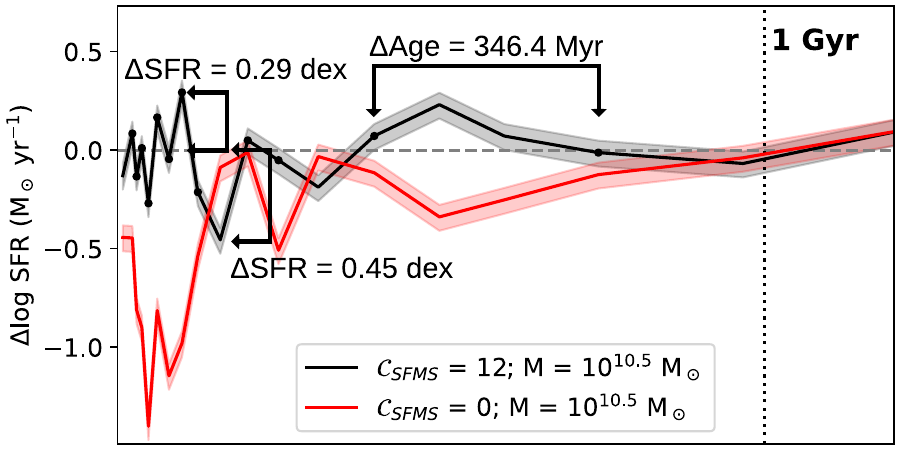}
\caption{Example of how crossings of the SFMS are detected. We show the position relative to the SFMS in dex of two galaxies with similar mass but very different number of crossings, obtained by subtracting the value of the SFMS at each age from the SFR at each age. The age bins where a crossing is detected (the previous age bin has opposite sign of $\Delta$log SFR) are indicated with a dot.}\label{fig:example}
\end{figure}

\begin{figure}[ht]%
\centering
\includegraphics[width=\linewidth]{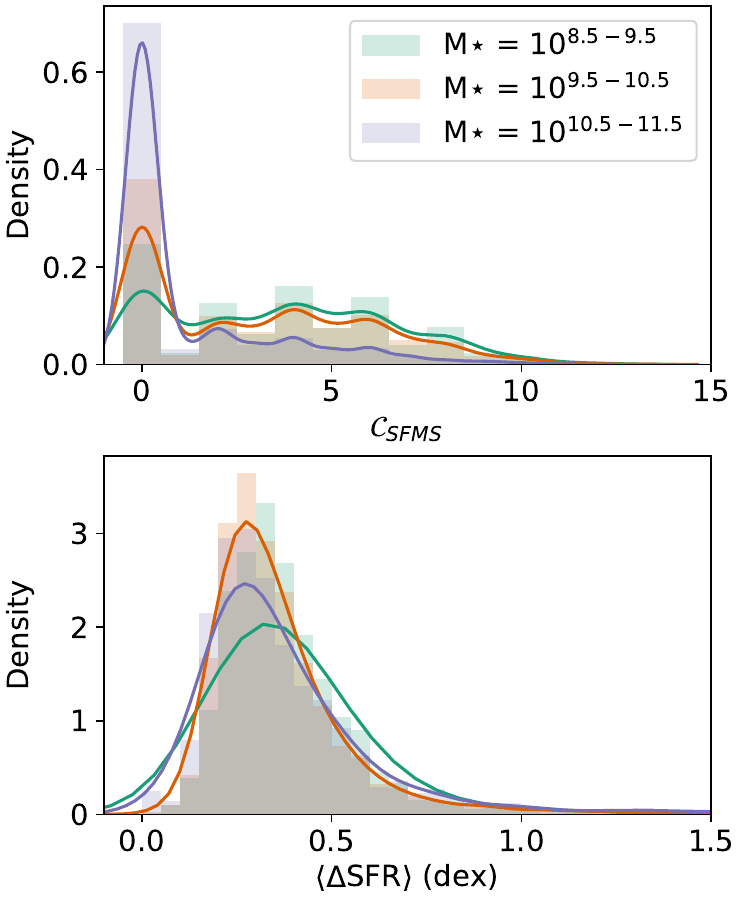}
\caption{Distribution of the parameters for ALL galaxies for three M$_\star$ bins. The top panel shows the average number of crossings per galaxy $\mathcal{C}_{SFMS}$ and in the bottom panel the average deviation from the SFMS $\Delta\mathrm{SFR}$ is shown. Each distribution is shown as both a histogram and the kernel density estimator (KDE) smoothed curve for each M$_\star$ bin. The kernel bandwidth is estimated using the default method from \texttt{SciPy}.} \label{fig:dist}
\end{figure}

Using the methodology outlined in the previous section we can obtain the age values at which each galaxy crosses the SFMS. We measure three parameters for each galaxy, (i) the total number of times it is crossed ($\mathcal{C}_\mathrm{SFMS}$), (ii) the average maximum deviation from the relation between two crossings measured in dex ($\Delta\mathrm{SFR}$), and (iii) the average time interval between two crossings ($\Delta$Age). These are proxies for the frequency of the variability, the intensity of the oscillations, and the duration of the deviation or period of oscillation, respectively. We consider both increases and decreases of the SFR relative to the SFMS, taking the absolute value of $\Delta\mathrm{SFR}$ when a galaxy goes below the relation. See Fig. \ref{fig:example} for an example indicating how the values are measured.

Because we measure the deviation relative to the SFMS, it is not reflective of how much a galaxy deviates from its steady state but instead how much it deviates from the average. This introduces a slight inconsistency in the case where galaxies have different long-term tracks. Consider a galaxy that is typically 0.1 dex below the SFMS and which has a burst of star-formation that increases its SFR by 0.2 dex. We would detect this burst as a pair of crossings of the SFMS but measure its $\Delta\mathrm{SFR}$ as 0.1 dex instead of 0.2 dex. Our deviations reflect how far from the average for its M$_\star$ a galaxy goes when it oscillates, rather than the intrinsic amplitude of its oscillations. This can also affect the measurement of the intervals, if the time it takes for a galaxy to reach the relation is longer than the age sampling.

In order to properly interpret the results shown here it is very important to understand how reliable the results obtained are. Given that we are measuring fluctuations in the values of the SFR and the abundance, we are uniquely affected by noise and random uncertainties which produce fluctuations in value as well. Because of this, we tested how the results would change if random noise was the origin of the scatter in SFR. The results, shown in Appendix \ref{sec:app-noise} prove that the distribution of values and the resulting correlations with M$_\star$ due to stochastic noise are distinct from those we measure. We are still very likely affected by random uncertainties, but not to the point that we cannot find underlying correlations.

In Fig. \ref{fig:dist} we show the distribution of values for the number of crossings and deviation from the SFMS for ALL galaxies, calculated as the average per galaxy and separated into three M$_\star$ bins. We do not include the interval $\Delta$Age due to how the uneven age sampling affects this measurement, we study this parameter in the following subsection. The distribution is shown using histograms as well as smoothed distributions calculated using kernel density estimation (KDE), which adaptively selects the smoothing parameters based on the data. We employ the \texttt{gaussian\_kde} method from \texttt{scipy} with the bandwidth estimation method from \cite{Scott2015}.

$\mathcal{C}_\mathrm{SFMS}$ shows zero as the most common number of crossings and less massive galaxies tend to have a higher number of crossings compared to more massive ones. Other than the peak at $\mathcal{C}_{SFMS}=0$ there is a wide distribution centered around $\mathcal{C}_\mathrm{SFMS}\sim4$ which constitutes most of the galaxies in the lowest M$_\star$ bin.
The distribution of $\mathcal{C}_\mathrm{SFMS}$ is very similar in shape for the two lower M$_\star$ bins in contrast to the most massive one, which shows a decline towards higher $\mathcal{C}_\mathrm{SFMS}$ rather than a wide peak. There is also a preference for an even number of crossings with about twice the fraction compared to and odd number of crossings, meaning that galaxies tend to deviate occasionally from their steady state before going back to it. We can infer that there is a certain value of SFR for each galaxy that represents its steady state, with intense bursts or quenching episodes occurring occasionally, as opposed to a scenario where they are constantly oscillating around a mean value. In the latter case there would be no preference for an even number of crossings, just as being noise-dominated would also produce a smooth distribution (see Appendix \ref{sec:app-noise}).
Regarding $\Delta\mathrm{SFR}$, the distribution becomes wider and shifts to higher deviations for the lower M$_\star$ bins but the distributions overlap almost completely.

\subsection{Measuring the time intervals} \label{sec:interval}
\begin{figure}%
\centering
\includegraphics[width=\linewidth]{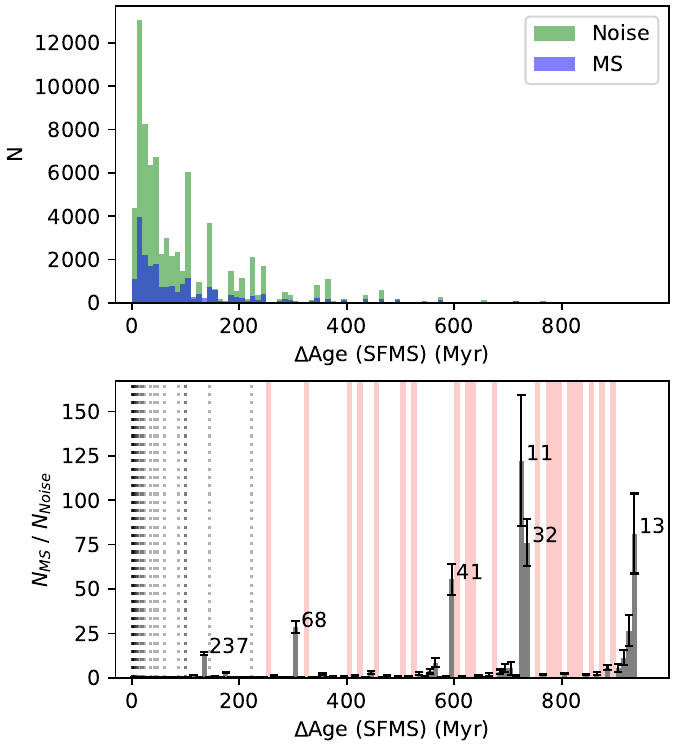}
\caption{Distribution of $\Delta$Age intervals for galaxies within the MS, sampled in $\Delta$Age bins 10 Myr wide. In the top panel our measured distribution is compared to the one resulting from imposing noise as the separation from the SFMS at each age value (see Appendix \ref{sec:app-noise}). In the bottom panel we show the ratio between the measured and noise-generated intervals for each bin. The peaks correspond to statistically overrepresented intervals in the measurements. Black bars show the error estimated assuming that the number of intervals measured follows a Poisson distribution.
Dotted lines show the age intervals between age bins within the last 1 Gyr and red shaded areas indicate the bins for which no combination of consecutive age bins is possible and are therefore impossible to be produced.}\label{fig:d_age_all}
\end{figure}

A third parameter to consider is the interval between crossings $\Delta$Age which can be taken as an approximate proxy for the period of the variability or the duration of bursts and/or small quenching episodes. Similar to $\Delta\mathrm{SFR}$, this measurement is affected by the offset between the typical position of a galaxy in the SFMS, as a galaxy needs to reach the average value for its M$_\star$ bin before its interval is measured. This offset can affect the results if the time it takes for an oscillating galaxy to reach the MS is longer than the sampling in age. Measuring the intervals using our dataset is particularly tricky due to the highly uneven sampling in age of our star formation histories (pseudo-logarithmic). Due to this, very short intervals can only be measured if they occur at a very recent age, with age steps ranging from a few Myr to $\sim200$ Myr. This makes it difficult to properly assess whether there are intrinsic time-scales associated with the variability of galaxies, as the age at which an oscillation occurs will play a significant role in determining the value of its interval.

To circumvent this problem, we have employed a statistical approach that will allow us to detect the presence of intrinsic time-scales and approximate their value. Assuming these time-scales exist, they would be overrepresented in the distribution of $\Delta$Age values compared to random fluctuations. Thus, we generated a new set of SFHs such that the distance to the SFMS for each galaxy at each age bin is randomly sampled from a normal distribution whose amplitude matches the scatter. For more details see Appendix \ref{sec:app-noise}. This new set of histories reproduces the SFMS at each age perfectly but erases any real, physical, information related to variability in the SFH. To estimate uncertainties and to improve the robustness of the calculation we have repeated the calculation for the noise-generated distribution 100 times and used the average at each $\Delta$Age bin. 

In Fig. \ref{fig:d_age_all} we show the distribution of $\Delta$Age resulting from our measurements as well as the one resulting from imposing random noise as the origin of the scatter in the SFMS. The general shape of the distributions is similar, but the real results have about a third of the crossings, showing that the detected fluctuations are mostly stochastic in nature but also that random uncertainties do not dominate our results. In the bottom panel of the figure, we show the ratio between the measured and noise-generated distributions. Any intrinsic time-scales that are more likely than would be expected from random fluctuations should show up here as peaks due to these $\Delta$Age values being overrepresented.

Indeed, in the ratio plots the larger peak centered at lower $\Delta$Age values disappears (comparatively) and a few very prominent peaks appear. Of particular relevance are the values obtained for the intrinsic time-scales in the SFMS: 135, 305, 595, 720-740, 900-940 Myr. These follow very closely a progression of multiples of $\sim$$150$ Myr: 150, 300, 450, 600, 750, 900 Myr with 450 Myr skipped, though there is a small peak at precisely this value.
The significance of the peaks is also very high, being between 10-120 times higher than the values expected for a noise-generated distribution (despite this distribution generating $\sim3$ times more crossings in general). It bears mention that the height of the peaks in the ratio between distributions is a measure of how significant the signal is but does not mean that the highest peak is the most likely interval to be observed. We show the number of crossings measured for each bin next to the peaks and it is clear that the 135 Myr peak is the most important time-scale.

The appearance of a linear sequence of $\Delta$Age$_\mathrm{SFMS}$ values is striking and strongly suggests the existence of an intrinsic time-scale for the variability of the SFMS of around $\sim$150 Myr. In this scenario the peaks at higher time-scales would result from the combination of two or more bursts/quenching episodes such that we do not measure the valley between them. For an example of sorts, see Fig. \ref{fig:example} at the location where $\Delta$SFR is indicated, this peak and the one to its left are detected as separate fluctuations, but if the data-point in between them was (due to noise or the sampling) a bit higher they would instead be detected as one fluctuation of double the interval.

The peaks in the ratio of the observed data to the simulated data in the left hand panels of Fig. \ref{fig:d_age_all} appear unnaturally sharp, however this is largely due to the use of evenly sampled histograms even though not all these bins can be populated equally or at all. The allowed values of $\Delta$Age are either the intervals of the sampling (indicated in the figure) or the sum of several consecutive intervals. This is the reason that, despite averaging 100 instances, the noise-generated distribution exhibits many peaks and null values (indicated as red zones in the figure) instead of a smooth distribution. Some of the bins have more opportunities to be populated and others are impossible to fill as there is no combination of intervals that lies in them. The expectation of a wider distribution around each peak is also not necessarily possible for all peaks. For each of the age bins within a $\Delta$Age interval there is a chance to artificially cross the SFMS due to noise which breaks up the interval into pieces of arbitrary length. For a regular sampling of the ages the adjacent ones would be expected to be populated more, but this is not our case. Populating the adjacent bins requires the lower age bins to be involved since their separation in age is smaller, and for the larger intervals, which span a large portion of the last 1 Gyr, the involvement of younger, smaller age bins is much more likely and that is why these do show wider distributions.
That being said, it is interesting to note that none of the prominent $\Delta$Age values coincide with the sampling, meaning that they must arise from a combination of age bins. Even the 135 Myr peak lies between sampling intervals, meaning that it is populated by oscillations that have several SFR measurements between the two crossings.

\subsection{Crossings vs M$_\star$}
\begin{figure}[ht]%
\centering
\includegraphics[width=\linewidth]{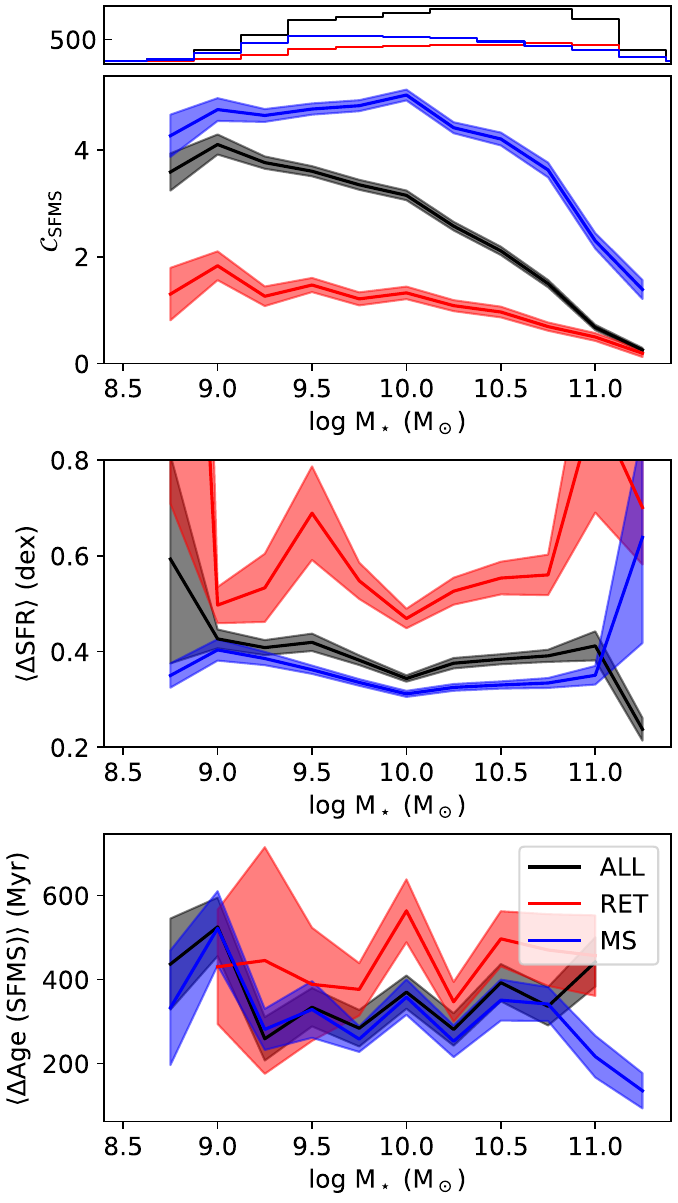}
\caption{Properties related to the crossings of the SFMS compared to M$_\star$, for the three groups of galaxies considered: ALL (black), MS (blue) and RET (red). The top panel shows the distribution of the galaxies in M$_\star$. In the second panel, the solid lines show the average number of crossings ($\mathcal{C}_\mathrm{SFMS}$) for each M$_\star$. In the third panel the solid lines show the $\Delta\mathrm{SFR}$ in dex after a crossing has happened. In the fourth panel the solid lines show the average interval ($\Delta$Age) between crossings for the galaxies, in Myr. For the bottom three panels, the shaded areas correspond to the error of the mean.} \label{fig:cross}
\end{figure}

In Fig. \ref{fig:cross} we show the correlation between M$_\star$ and the parameters of the crossings for the three selected groups described in Sec. \ref{sec:cross_det}: ALL, MS and RET. For $\mathcal{C}_\mathrm{SFMS}$ and $\Delta\mathrm{SFR}$ we averaged the values for each galaxy within 0.25 dex wide M$_\star$ bins and the errors correspond to the error of the mean (scatter divided by $\sqrt{N}$) computed within each bin. For the $\Delta$Age values we follow a similar approach to the previous section. At each bin, we calculate the distribution of $\Delta$Age values for the galaxies included in it as well as the distribution if those same galaxies had noise-like offsets from the SFMS over time (see Appendix \ref{sec:app-noise}). We then produce the ratio between the distributions and select the peaks where the S/N reaches above 10. Each selected $\Delta$Age value is included in a weighed sum whose weights are how many times each value is measured. In this manner we remove those $\Delta$Age values that cannot statistically be distinguished from random fluctuations and then obtain a representative average $\mu$ such that the more a $\Delta$Age value appears the more weight it has on the average. For the error we estimate the variance using the second moment with the same weights:
\begin{equation}
    \mu = \frac{\sum_i N_i\cdot\Delta\mathrm{Age}_i}{\sum_i N_i}
\end{equation}
\begin{equation}
    \sigma^2 = \frac{\sum_i N_i\cdot(\Delta\mathrm{Age}_i-\mu)^2}{\sum_i N_i}
\end{equation}
The value of $\sigma$ is then used to calculate the error of the mean dividing it by the square root of the total number of $\Delta$Age values for the selected peaks with S/N above 10 ($\sqrt{\sum_i N_i}$).

In Fig. \ref{fig:cross} the number of crossings of the SFMS strongly varies between MS and RET galaxies, the former of which show a fairly flat $\sim$$4.5$ average crossings below \logm 10 at which point it starts to decline. RET galaxies, on the other hand, have significantly fewer crossings on average for all masses ($\sim$$1-2$) and the number of crossings declines steadily with mass from 2 to almost zero for the most massive galaxies. The average $\Delta$SFR is fairly constant for MS but varies irregularly for RET, likely due to the lower number of both galaxies and average number of crossings for this bin. Still, there is a clear offset compared to MS such that the deviations are higher.

Since RET galaxies have on average $\mathcal{C}_{SFMS}\lesssim2$ and they are selected to currently be 1$\sigma$ below the SFMS, a significant portion of this sample are galaxies which have crossed the SFMS one last time within the last 1 Gyr and are currently on their way to becoming retired. In Fig. \ref{fig:example} the RET galaxy almost reaches the SFMS one last time before its SFR drops below -1.25 dex at the lowest point, a much larger offset than the typical deviations shown by the MS galaxy.

The averaged values for $\Delta$Age$_\mathrm{SFMS}$ do not appear to change significantly with M$_\star$, other than perhaps having slightly higher values at $10^9$ M$_\odot$ and below and shorter time-scales above $10^{10.75}$ M$_\odot$. An offset compared to RET can also be seen, but these features are not very clear compared to the uncertainties so in general we do not detect strong trends in this parameter with M$_\star$ or between MS and RET.

In Fig. \ref{fig:cross} the average values for $\Delta_\mathrm{SFR}$ show an offset between RET and MS galaxies, but we do not observe two peaks in the global distribution in Fig. \ref{fig:dist} despite the difference in value (0.4 to 0.6) being significant enough that it would be detectable. The contribution of RET galaxies to the distribution in Fig. \ref{fig:dist} is therefore the extended wing towards higher $\Delta$SFR that makes the distribution asymmetric. For MS galaxies their average values coincide with the peak in the distribution at $\sim$0.4 dex, suggesting that the intensity of a burst of star-formation or a quenching episode for a galaxy on the MS is, on average, a fundamental value $\gtrsim0.4$ dex. This means that, regardless of their mass, when galaxies have a burst of star-formation or a quenching episode they typically vary their SFR by $\sim$2.5 times. This value is lower than typically reported offsets of the starburst population of $0.6-1$ dex \citep{Rodighiero2011,Sargent2012, Rodighiero2014, Schreiber2015}, but we need to consider that we require a crossing of the SFMS to occur before we measure the deviation. As such, if we use $\Delta_\mathrm{SFR}$ as a proxy for the position of the starbursts, we are selecting galaxies which typically lie below the SFMS and ignoring those that typically lie above it (as the latter will not cross despite the increase). This biases $\Delta_\mathrm{SFR}$ to lower values by a factor that should be of order $\sim$$ 0.5\cdot\sigma_{SFMS}$ which given $\sigma_{SFMS} = 0.4$ makes a total $\Delta$SFR of 0.6 matching the lower end of the expected values. Also, $\Delta_\mathrm{SFR}$ includes the deviation from galaxies temporarily dipping below the SFMS, which might have a lower amplitude for the MS sub-sample. Finally, the SSP sampling can effectively "smooth" the peaks in the SFH compared to emission line measurements. Because of this, we consider our results to be roughly compatible with the reported offset for star-bursts accounting for a systemic offset of 0.2 dex due to how it is measured. Therefore, the narrow distribution of $\Delta_\mathrm{SFR}$ that does not depend on M$_\star$ indicates that individual bursts of star-formation occur in a similar, self-regulated manner, in all galaxies with a similar relative increase in SFR whose value is not strongly affected by halo-level parameters.

In general, we can characterize the main results for the crossings as such: (i) more massive galaxies have fewer crossings in general no matter what, (ii) MS galaxies have more crossings in general, (iii) $\Delta\mathrm{SFR}$ is fairly constant with M$_\star$ but offset to larger values for RET galaxies, (iv) we find no clear correlation between $\Delta$Age and M$_\star$.
These results support star-formation bursts or quenching episodes as self-regulated incidents where the relative intensity and the duration do not depend on the global parameters of the host galaxy, whose role is to instead regulate whether and how often the bursts occur. 

\subsection{Crossings vs current position in the SFMS and MZR} \label{sec:cross_position}

\begin{figure}%
\centering
\includegraphics[width=\linewidth]{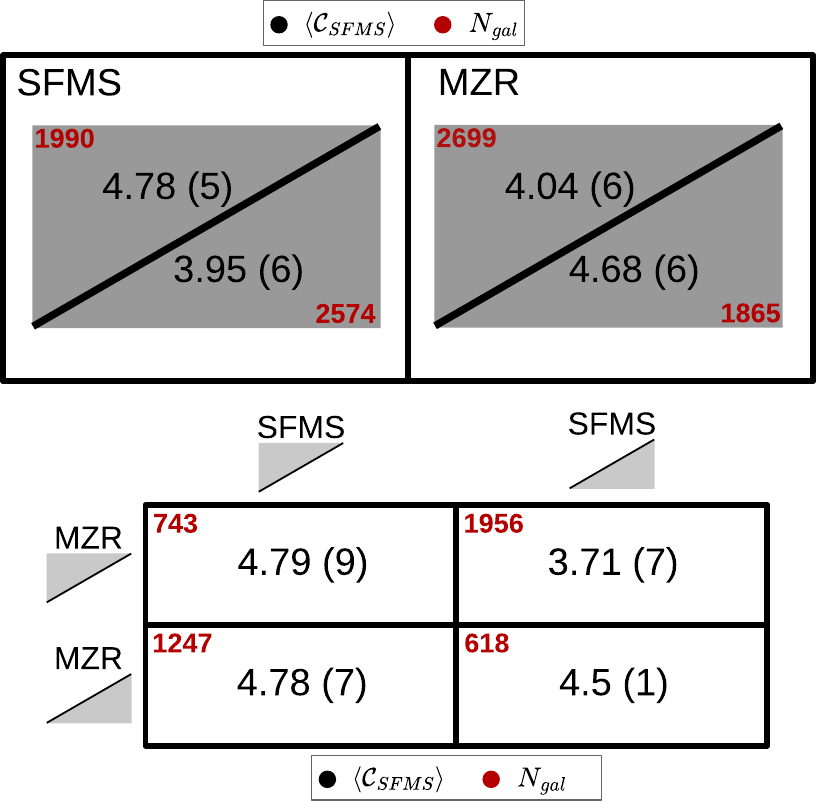}
\caption{Average number of crossings ($\mathcal{C}_\mathrm{SFMS}$) for several bins of galaxies which are selected based on the galaxies' currently observed position in the MZR and SFMS. In the upper panel, the diagonal lines represent the relations so that the position of the numbers relative to them informs which positional bin (above or below the relation) is being considered. In the bottom panel the bins considering combinations of positions in both relations are shown, with the shaded areas indicating the position in each relation.
The error is computed as the error of the mean and indicated in parenthesis as first significant digit in the error corresponding to the last digit shown in the value. In the corners in red we indicate the number of galaxies included in each bin.}\label{fig:tab_cross}
\end{figure}

\begin{figure}%
\centering
\includegraphics[width=\linewidth]{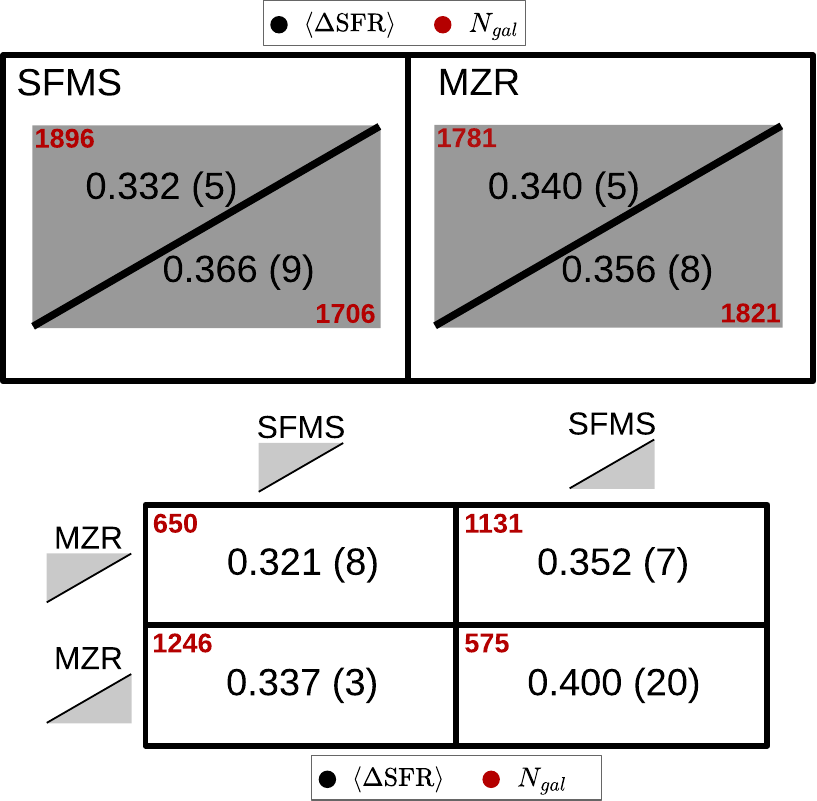}
\caption{Same as Fig. \ref{fig:tab_cross} but for the deviation from the SFMS ($\Delta\mathrm{SFR}$). In order for galaxies to be included in a particular bin they need to have $\mathcal{C}_\mathrm{SFMS}> 0$ and therefore the number of galaxies per bin is different from that in Fig.\ref{fig:tab_cross}.}\label{fig:tab_dev}
\end{figure}

\begin{figure}%
\centering
\includegraphics[width=\linewidth]{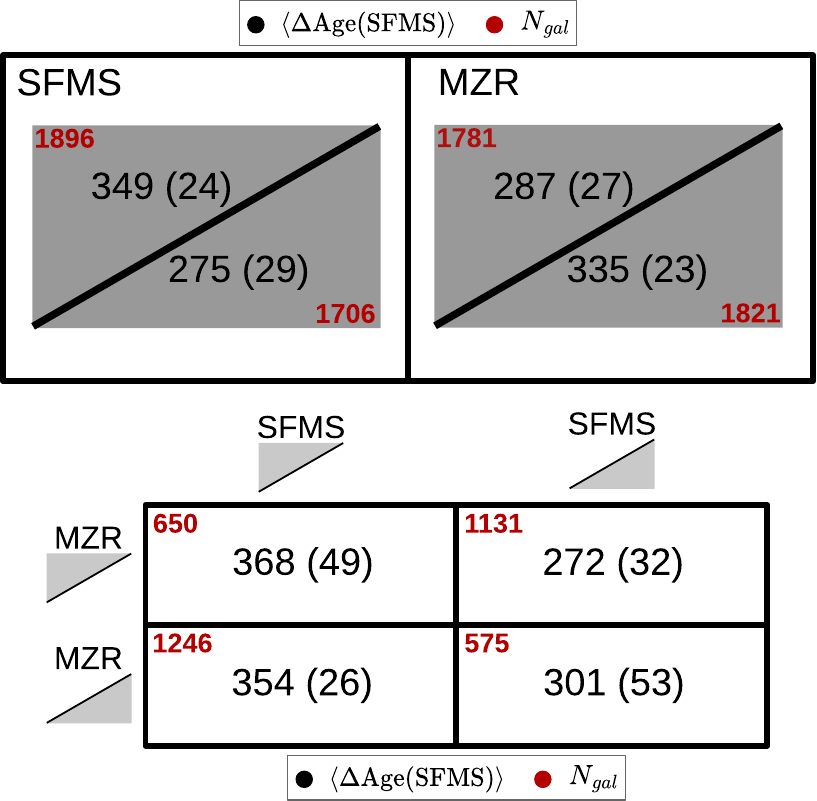}
\caption{Same as Fig. \ref{fig:tab_cross} but for the interval (in Myr) between crossings from the SFMS. In order for galaxies to be included in a particular bin they need to have $\mathcal{C}_\mathrm{SFMS}> 0$ and therefore the number of galaxies per bin is different from that in Fig.\ref{fig:tab_cross}.}\label{fig:tab_int}
\end{figure}

Another way to probe how the number of crossings can be tied to the evolution of the galaxies is to check how the averaged parameters vary depending on where the galaxies are currently located in the MZR and SFMS. In other words, we measure the average parameters for galaxies that are currently above or below either the MZR or SFMS. For this analysis we employ only the MS group of galaxies, as the RET galaxies cannot be currently located above the SFMS by definition. As can be seen in the figures, the number of galaxies above and below the relations is not exactly the same even selecting only MS galaxies. This is because the SFMS and MZR are calculated using galaxies selected with EW$_\mathrm{H\alpha} > 6$\AA{}, while the MS sample requires galaxies to be within $\pm 1 \sigma$ of the SFMS. In theory, this should be roughly equivalent to selecting MS after determining the SFMS but, as with all empirical relations, there is scatter such that some galaxies are currently measured as within the MS using full spectral fitting but have lower EW$_\mathrm{H\alpha}$ so they're not selected for measuring the SFMS. The result is an excess of galaxies below the MS and above the MZR, the latter is due to how galaxies with lower SFR are located above the MZR (see Fig. \ref{fig:current}). The difference in galaxies for the SFMS is about 12\% of the MS sample.

In Figs. \ref{fig:tab_cross}, \ref{fig:tab_dev}, and \ref{fig:tab_int} we show the results for the three parameters,  $\mathcal{C}_\mathrm{SFMS}$, $\Delta\mathrm{SFR}$, and $\Delta$Age, respectively. The position of the numbers in the diagrams indicates which bin of galaxies is considered. In general, the differences between bins are smaller than those observed for M$_\star$ and some parameters appear largely unaffected by the current position of the galaxies. One thing to consider when interpreting these results is that, ideally, we would select galaxies based on their "preferred" location within the relations rather than their current one which could be due to a temporary oscillation.

One of the most relevant correlations appears to be in $\mathcal{C}_\mathrm{SFMS}$, which is significantly higher for galaxies that are either currently above the SFMS (>SFMS) or below the MZR (<MZR) when considering only one relation. Considering the position in both relations at once, this difference appears to be driven by the bin located at $<$SFMS \& $>$MZR, which has the lowest $\mathcal{C}_\mathrm{SFMS}$. This result fits well with pristine gas accretion sustaining star formation in galaxies, such that those that are deficient in gas accretion will not be able to fuel as many bursts of star-formation. Since the gas has very low abundances, galaxies that accrete less of it will also tend to be located at higher average abundances for their M$_\star$. We do not observe the opposite where the $>$SFMS \& $<$MZR bin has a higher average $\mathcal{C}_\mathrm{SFMS}$, however, which might indicate that changes in SFR and [Z/H] are not simultaneous in general, allowing us to find galaxies with large gas accretion in any combination of positions of the SFMS and MZR. The accretion deficient galaxies, on the other hand, do not shift around the SFMS as much and therefore correlate with their current position in both relations much more.

For $\Delta\mathrm{SFR}$ we find higher deviations for the bins below the SFMS and MZR, but the differences are much smaller relative to the values of $\Delta\mathrm{SFR}$ and closer to compatibility with the error. $\Delta$Age follows similar correlations to $\mathcal{C}_\mathrm{SFMS}$, with galaxies $<$SFMS \& $>$MZR having longer intervals.

Another interesting parameter to compare is the number of galaxies per bin. The distribution of galaxies is uneven, such that most galaxies lie on opposite sides of both relations, that is, either $>$SFMS \& $<$MZR bin or vice-versa. This suggests a co-evolution of both relations such as the fundamental metallicity relation \citep{Mannucci2010}. The number of galaxies per bin changes for Figs. \ref{fig:tab_dev} and \ref{fig:tab_int}, as galaxies with no crossings do not contribute to measuring the deviation and interval. The greatest difference in the number of galaxies appears for the aforementioned bin that is expected to be deficient in gas accretion ($<$SFMS \& $>$MZR) with 42\% of the galaxies showing no crossings of the SFMS. Strikingly, in the bin opposite to it ($>$SFMS \& $<$MZR) only one galaxy out of 1247 shows no crossings, and for the other two bins the fraction is 7\% and 13\%. This result agrees very well with the above hypothesis that the $<$SFMS \& $>$MZR bin is the location for galaxies deficient in gas accretion. Since these galaxies still belong to the MS sample it is likely that they are just now starting to retire and will soon cross into the Green Valley. The fact that the position in the MZR is the determining factor to detect them shows that, as proposed in \cite{Camps-Farina2023}, galaxy quenching generally occurs after the ISM stops being significantly diluted, likely due to gas accretion being shut off.

\subsection{Fraction of time above the SFMS} \label{sec:time}
\begin{figure}[ht]%
\centering
\includegraphics[width=\linewidth]{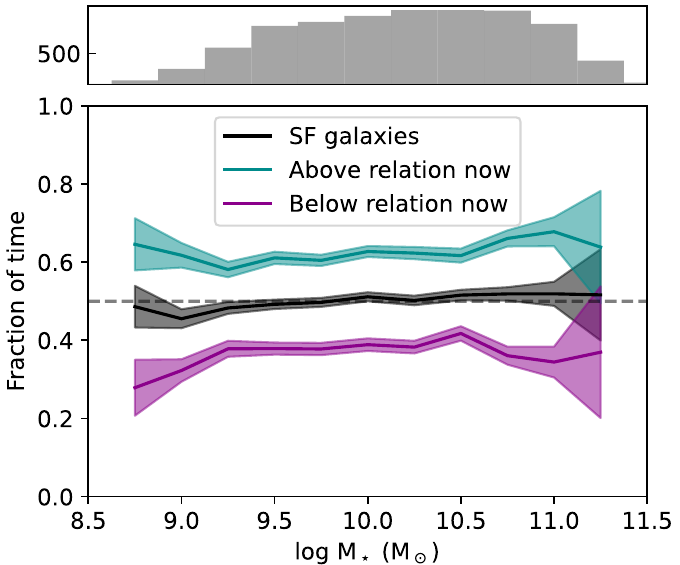}
\caption{Fraction of time spent above the SFMS in the last 1 Gyr for currently star-forming galaxies. The shaded areas correspond to the error of the mean.} \label{fig:perc}
\end{figure}

So far, we have characterized the motions of the galaxies around the SFMS, showing that at least some of its scatter originates in short-term (several to hundreds of Myr) variability in SFR. The question remains whether these fluctuations can account for most if not all the scatter. If they do not, it means that galaxies of the same mass have different average values of the SFR over time. It would result in some galaxies being intrinsically more star-forming than others. These different evolutionary tracks are expected to come from differences in halo-level properties such as halo formation time \citep{Matthee2019,Blank2021}.
In order to assess this, we have measured the percentage of time that star-forming galaxies currently above and below the SFMS have spent above it.

In Fig. \ref{fig:perc} we show the fraction of time above the SFMS vs M$_\star$ for star-forming galaxies selected using the EW$_\mathrm{H\alpha}$ > 6 \AA{} criterion (see Sec. \ref{sec:cross_det}). As explained in Sec. \ref{sec:cross_position}, the MS sample includes 12\% more galaxies that do not fulfill the EW$_\mathrm{H\alpha}$ criterion used to determine the SFMS. These are preferentially located below the SFMS and are likely galaxies whose SFR is beginning to decline. This makes them less suitable for the purpose of this section, which is checking whether actively star-forming galaxies tend to stay on one side of the relation. As such, instead of the MS sample we apply the EW$_\mathrm{H\alpha}$ criterion which is independent of the full spectral fitting and therefore makes for more robust results. This sample contains 4253 galaxies.

We show only the fraction of time above the SFMS and not that of time below it because these are interchangeable, and we have ensured that for all galaxies both fractions add up to 1.
There is a clear offset in the fraction such that, while the sample as a whole averages to galaxies spending half their time above the SFMS as would be expected, galaxies currently located above or below the relation tend to have spent, on average, most of their time on the same side ($\sim$$60$\%).

This result shows that a significant part of the scatter of the SFMS originates in galaxies having intrinsically different values of SFR at the same M$_\star$. This could be due to either a difference in the efficiency at which molecular gas is converted into stars (SFR/M$_\mathrm{H2}$), the same but for HI to H$_2$ conversion, a difference in the gas fraction of the galaxies, or a difference in the efficiency at which the gas is condensed into molecular gas. As mentioned above, differences in halo-formation time are a likely explanation for this though in that case a better interpretation of the scatter of the SFMS is differences in M$_\star$ originating in some halos having had less time than others to form stars \citep{Matthee2019,Blank2021}. In other words, two galaxies with the same SFH would always have similar SFR but if we offset them in time, due to one dark matter halo forming earlier, then for a typical t-tau shape which declines after cosmic noon there will be an offset between them such that the one that formed later will have higher SFR.
Environment is also a likely suspect as it can affect most of the mechanisms listed above and acts on long time-scales \citep[e.g.,][]{McGee2009,Haines2015,Paccagnella2016,Maier2019,Kim2023}. The first explanation could be checked by using the halo-mass instead of M$_\star$ for Fig. \ref{fig:perc}, which should bring the fractions closer to 0.5.

\section{Discussion}\label{sec:discussion}
Full spectral fitting with stellar populations is rife with degeneracies and systematic uncertainties \citep[e.g.][]{Walcher2011, CidFernandes2014}. This makes quantitative determinations of individual galactic parameters less reliable, but they work well for statistical comparisons between galaxies such as the results we present here. The actual number of crossings and intervals shown in each figure are heavily model dependent, primarily on the number of age values present in the SSP library employed. These represent the number of times a galaxy can be detected to have crossed the SFMS and therefore we are largely blind to crossings which occur with higher frequency. Individual histories can be rather noisy too and these surely contribute to the total number of crossings detected on each galaxy. The reliable results are therefore the statistically significant differences between bins of galaxies that we report.

Perhaps one of the most surprising results is the detection in Sec. \ref{sec:interval} of what appears to be a statistically preferred time-scale of $\sim$135-150 for SFR fluctuations such that this interval and its multiples appear much more frequently than would be expected for random fluctuations. The fraction of crossings associated to the $\sim$$135-150$ Myr time-scale and its multiples is quite small ($\sim$$400$ out of $\sim$$22,000$), but it has very high statistical significance. This possibly points to either the existence of a regulatory mechanism that acts on this time-scale or for this time-scale to be intrinsic to the triggering and propagation of star-formation over a galaxy, such as a crossing time for self-propagated star formation. This feature must arise from either real physics or from systematic method-related uncertainties that correlate with this particular time-scale of 135-150 Myr. It is never wise to assert that systematic effects are not present in any measurement obtained using full spectral fitting techniques, as they are numerous and difficult to fully disentangle from the results. That being said, our age sampling is pseudo-logarithmic in nature, and changes in the shape of the SSP templates do not generally follow linear progressions such as the one seen here. We are not aware of any systematic effect capable of artificially producing this feature, especially since the values are not perfect multiples of each other, which would be expected of a systematic, method-induced, intrinsic time-scale. We consider this a promising result, but proper verification is required from other authors using different SSP templates and fitting algorithms for its detection to be certain. 

From Fig. \ref{fig:d_age_all} it might be tempting to dismiss most of the detected crossings as noise, since the general shape of the distribution vanishes when divided by a noise-like distribution. However, this is not actually an inference that can be made from this figure. What we are measuring is how distinct from random fluctuations the distribution is, which allows us to detect the presence of the overrepresented $\Delta$Age values we report. However, just because they follow a random-like distribution does not mean they originate from uncertainties, as they can be the result of physical stochastic processes that would produce the same distribution. In other words, the similarity between the two distributions can also be interpreted as the triggering of star-formation being mostly a stochastic process. In Sec. \ref{sec:time} we show clearly that galaxies have a preferred side of the SFMS in which to reside, which we would not be able to measure if we were dominated by random uncertainties. The much lower number of crossings for the MS compared to the noise-like distribution in Fig. \ref{fig:d_age_all} ($\sim$$1/3$) also reflects this. The best evidence that the smaller $\Delta$Age values are not dominated by noise is in Fig. \ref{fig:dist}, which shows that the distribution of crossings is strongly biased to even numbers. In Fig. \ref{fig:app-dist} no such feature can be seen in the noise-like distribution of $\mathcal{C}_\mathrm{SFMS}$.

In conclusion, our study of the distribution of $\Delta$Age$_\mathrm{SFMS}$ values points to a scenario where star formation is triggered stochastically in sub-100 Myr time-scales, but these are modulated by longer bursts of star-formation regulated to last about 135-150 Myr. This scenario of larger bursts of star-formation of scale $\sim$100 Myr within which there are several shorter, $\sim$10 Myr individual bursts fits well with previous studies \citep{ForsterSchreiber2003,DiMatteo2008,McQuinn2009,French2018}, lending credibility to our results. 

One of the main questions we wanted to answer is whether short-term oscillations can explain the scatter of the SFMS. We find that the average deviation for MS galaxies is of a similar magnitude to the scatter of the relation (0.3-0.4 dex), but we also show that galaxies currently above and below the SFMS tend to spend most of their time on the same side for the past Gyr. In our results the excess is moderate, with a 60-40\% split rather than the 50-50\% that would arise from purely stochastic fluctuations, but as shown in Fig. \ref{fig:app-perc2} uncertainties strongly affect this value bringing it closer to 50-50. The 60-40\% split is reproduced when noise equal to $\sigma_\mathrm{SFMS}$ is introduced to SFHs which have no intrinsic fluctuations and their offset reproduces the scatter, so the offset between SFHs is likely a significant contributor to $\sigma_\mathrm{SFMS}$ given that we are strongly affected by noise.
Consequently, our results favor the mixed scenario predicted by \cite{Matthee2019, Tacchella2020} where galaxies are mostly spread around the SFMS because of long-term variation star-formation and short-term variability contributes to increasing the scatter. In our results, it interesting that the number of crossings declines for high mass galaxies but the scatter of the SFMS increases, going from 0.34 at \logm 9 to 0.56 at \logm 11. Since the scatter increases even though the oscillations become less frequent it should mean that the long time-scale differentiations in the SFH account for a larger portion of the scatter of the SFMS in more massive galaxies. Oddly, such a M$_\star$ dependence is not observed in Fig. \ref{fig:perc} for the SFMS.
The dependence of the scatter of the SFMS with M$_\star$ is still debated, several observational studies \citep{Brinchmann2004, Guo2013, Popesso2019} find an increase with mass, especially a jump in scatter at masses approaching \logm11 just like us, and \cite{Sparre2015} finds the same in the Illustris simulations. However, \cite{Matthee2019} finds the opposite in EAGLE simulations in agreement with the observations of \cite{Chang2015}.

In Fig. \ref{fig:current} we show where MS and RET galaxies are located in both relation highlighting that, while MS galaxies occupy most of the parameter space that the full sample covers, RET galaxies are strongly shifted to higher metallicity values. This fits very well with the expectation that pristine gas accretion is required for galaxies to stay in the MS \citep{Camps-Farina2023}. \cite{Vaughan2022}, however, finds that this offset in the MZR between star-forming and passive galaxies correlates with the size of the galaxies in observational data, such that due to evolutionary tracks smaller galaxies tend to be the most metallic for their M$_\star$.

\section{Summary and Conclusions}\label{sec:conclusions}
In this article we have used full spectral fitting to obtain the SFH and [Z/H] for 8960 MaNGA galaxies so that we can study how the SFR changes over time and check whether the scatter of the SFMS can be attributed to short-term fluctuations or if instead galaxies have offsets in these parameters that are sustained over time. Because of how close to the uncertainties our measurements lie, we need robust statistical measurements to make sure that our results are significant, which we have achieved mostly by making sure that our results cannot arise from random uncertainties and that they do not follow expected trends arising from fitting degeneracies.

In general, we find that this type of study is feasible with large datasets and find results that fit well with what is known in the field and predictions from simulations, making this a promising methodology especially for application to future spectroscopic surveys of galaxies with improved spectral resolution and S/N.

Some of the most important results are:
   \begin{enumerate}
      \item SFHs obtained from full spectral fitting are capable of yielding useful results on short time-scales. We use this to measure the variability in SFR by measuring how galaxies cross the SFMS over the last 1 Gyr.
      \item Variability in the SFH appears to be a combination of purely stochastic sub-100 Myr fluctuations and longer bursts or quenching episodes with a possible slight excess on a 135-150 Myr time-scale and its multiples.
      \item The number of crossings of the SFMS ($\sim4.5$) does not appear to change much with M$_\star$ in the 10$^{8.5-10.5}$ M$_\odot$ range, above which galaxies tend to fluctuate less.
      \item The degree to which galaxies deviate from the SFMS does not appear to correlate strongly with M$_\star$ and is similar to the scatter of the relation at 0.3-0.4 dex.
      \item The current location in the SFMS and the MZR has a significant effect on how galaxies cross the former, pointing to a tie between the enrichment history and the SFH. Galaxies located below the SFMS and above the MZR show the lowest number of crossings, likely because 42\% of galaxies located here have not had a burst of star-formation in the last 1 Gyr. These galaxies are likely to fall to the Green Valley soon.
      \item Galaxies appear to have preferred locations in the SFMS such that they spend 60\% of the last Gyr there. This shows that the scatter of the SFMS is a combination of long- ($>1$ Gyr) and short-term ($\lesssim$100 Myr) fluctuations, in agreement with models.
   \end{enumerate}

\begin{acknowledgements}
We thank the referee for their useful comments and S. Pascual for discussion on the analysis.
ACF acknowledges financial support by the Spanish Ministry of Science and Innovation through the research grants PID2019-107427-GB-31 and PID2022-138855NB-C31, funded by MCIN/AEI/10.13039/501100011033/FEDER, EU. SFS thanks the PAPIIT-DGAPA AG100622 project and CONACYT grant CF19-39578. This work was supported by UNAM PASPA – DGAPA. MCC acknowledges the support of AC3, a project funded by the European Union's Horizon Europe Research and Innovation programme under grant agreement No 101093129, PID2022-138621NB-I00, funded by MCIN/AEI/10.13039/501100011033/FEDER, EU. This work has been partially funded by "Tecnolog\'ias Avanzadas para la exploraci\'on del universo y sus componentes" (PR47/21 TAU-CM) project funded by Comunidad de Madrid, by the Recovery, Transformation and Resilience Plan from the Spanish State, and by NextGenerationEU from the European Union through the Recovery and Resilience Facility.

Funding for the Sloan Digital Sky Survey IV has been provided by the Alfred P. Sloan Foundation, the U.S. Department of Energy Office of Science, and the Participating Institutions. SDSS acknowledges support and resources from the Center for High-Performance Computing at the University of Utah. The SDSS web site is \url{www.sdss.org}.

SDSS is managed by the Astrophysical Research Consortium for the Participating Institutions of the SDSS Collaboration including the Brazilian Participation Group, the Carnegie Institution for Science, Carnegie Mellon University, Center for Astrophysics $\vert$ Harvard \& Smithsonian (CfA), the Chilean Participation Group, the French Participation Group, Instituto de Astrof\'{i}sica de Canarias, The Johns Hopkins University, Kavli Institute for the Physics and Mathematics of the Universe (IPMU) / University of Tokyo, the Korean Participation Group, Lawrence Berkeley National Laboratory, Leibniz Institut f\"{u}r Astrophysik Potsdam (AIP), Max-Planck-Institut f\"{u}r Astronomie (MPIA Heidelberg), Max-Planck-Institut f\"{u}r Astrophysik (MPA Garching), Max-Planck-Institut f\"{u}r Extraterrestrische Physik (MPE), National Astronomical Observatories of China, New Mexico State University, New York University, University of Notre Dame, Observat\'{o}rio Nacional / MCTI, The Ohio State University, Pennsylvania State University, Shanghai Astronomical Observatory, United Kingdom Participation Group, Universidad Nacional Aut\'{o}noma de M\'{e}xico, University of Arizona, University of Colorado Boulder, University of Oxford, University of Portsmouth, University of Utah, University of Virginia, University of Washington, University of Wisconsin, Vanderbilt University, and Yale University.

\end{acknowledgements}

\bibliographystyle{aa} % style aa.bst
\bibliography{export-bibtex.bib} % your references Yourfile.bib

% \end{document}

\begin{appendix}%First appendix

\section{Primary and secondary samples in MaNGA}\label{sec:app-primsec}
The MaNGA sample is composed of 3 sub-samples: Primary, Secondary and Color-enhanced. The first two account for most of the galaxies (84\%) and the Color-enhanced one was added to increase statistical significance in areas of the NUV-i vs M$_{i}$ parameter space that are more sparsely populated.
The Primary and Secondary samples are defined such that at least 80\% of the galaxies contained in them are covered by the IFU out to 1.5 and 2.5 times the effective radius, respectively. Consequently, our SFR and M$_\star$ determinations for the Primary sample are measured using a lower fraction of light compared to the the Secondary sample. The metallicity values are less affected by this due to being measured at the effective radius.

\begin{figure}%
\centering
\includegraphics[width=\linewidth]{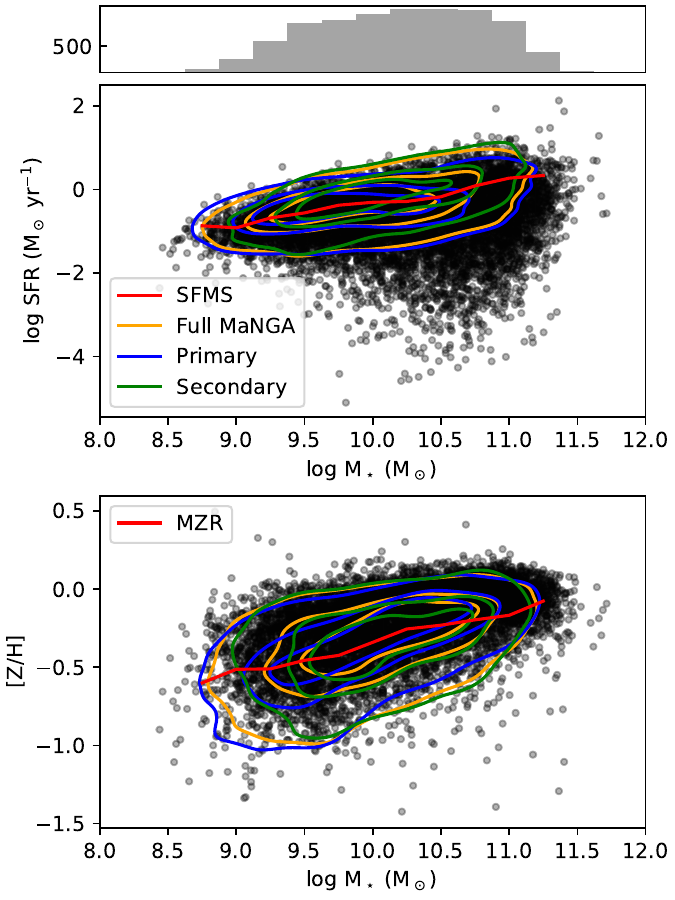}
\caption{Comparison between the Primary and Secondary samples in MaNGA regarding their distribution in the SFMS (top) and MZR (bottom) relations. In both panels the black points correspond to all galaxies in the sample, while the red lines (averaged SFMS and MZR) and contours (35\%, 65\% and 95\%) represent the EW$_\mathrm{H\alpha}$ selected star-forming galaxies used to measure the relations. For the orange contours no further selection based on MaNGA target selection, while the blue and green contours select, respectively, the Primary and Secondary samples.}\label{fig:app-samples_current}
\end{figure}

The variability of the SFR should not be strongly affected by this except in galaxies where the bulk of star-formation occurs at the outskirts, which should be relatively unusual. That being said, if there is a significant statistical difference in the total SFR measured this could manifest as an offset between the samples in the SFMS such that galaxies in the Secondary sample lie preferentially above or below it. In Fig. \ref{fig:app-samples_current} we show the SFMS and MZR as in Fig. \ref{fig:current} but with contours for the Primary and Secondary samples, with no significant offset between the samples observed.

\section{How distinct are our results from pure noise?} \label{sec:app-noise}
There is a variety of ways in which our results are affected by the method, chiefly among them the sampling of the stellar population ages and the intrinsic uncertainties in the determination of the parameters. Additionally, the latter of the two are not well constrained. As such, it is of interest to check how our results change if we impose random noise as a dominant factor, to see if we find significantly distinct values.

\subsection{Change in our parameters if crossings are dominated by noise}

\begin{figure}[ht]%
\centering
\includegraphics[width=\linewidth]{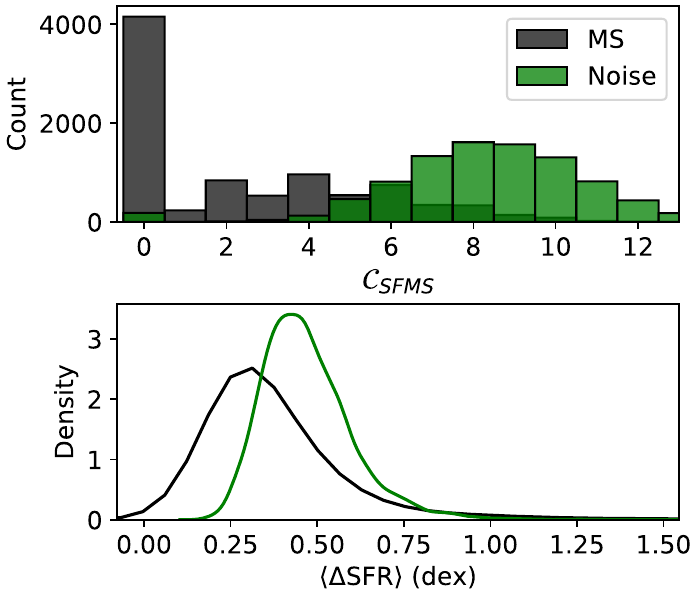}
\caption{In the top panel the distribution of values for $\mathcal{C}_\mathrm{SFMS}$ is shown, while the $\Delta$SFR distributions are shown in the bottom panel. The actual measured values are shown in black while the predicted values from assuming purely random fluctuations as the origin of the scatter are shown in green.} \label{fig:app-dist}
\end{figure}

\begin{figure}[ht]%
\centering
\includegraphics[width=\linewidth]{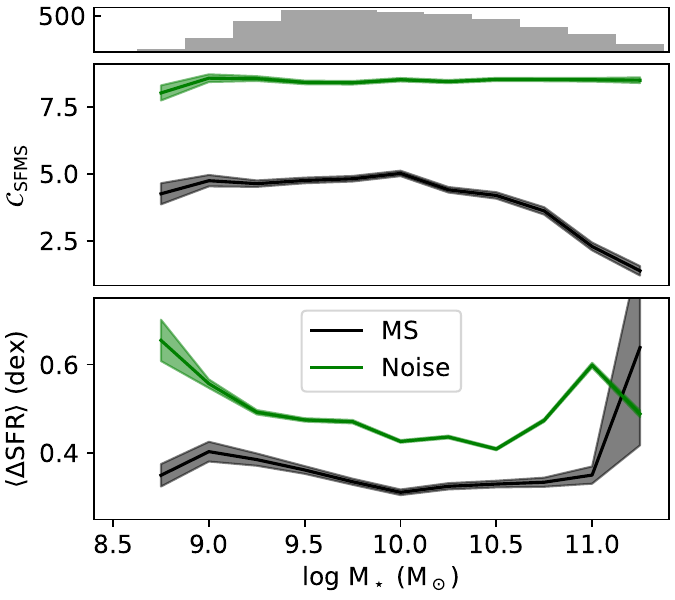}
\caption{Properties related to the crossings of the SFMS compared to M$_\star$. The actual measured values for MS galaxies are shown in black while the predicted values from assuming purely random fluctuations as the origin of the scatter are shown in green.} \label{fig:app-cross}
\end{figure}

If we substitute the residuals of the SFMS we use to detect crossings (see Sec. \ref{sec:cross_det}) with random values sampled from a Gaussian distribution of width equal to the measured width of the SFMS at the corresponding M$_\star$ bin we reproduce the overall distribution of the relation but erase any physical information in the galaxy histories.
The specific age sampling will produce characteristic average values for each of the parameters when processing noise. For example, we would expect the average number of crossings to be close to half the number of age values since there is a 50\% chance for the galaxy to switch sides on the SFMS if its position is determined randomly.

In Fig. \ref{fig:app-dist} we show the distribution for two of the parameters of the crossings for MS galaxies as well as what results from imposing noise as the origin of the scatter. The methodology to measure representative values for $\Delta$Age shown in Sec. \ref{sec:interval} already includes a comparison to the noise-like distribution and therefore cannot be included here.
For $\mathcal{C}_\mathrm{SFMS}$ and the $\Delta$SFR the distributions for MS and noise are different, though the position of the peaks are somewhat similar in $\Delta$SFR. The average number of crossings, which is the most important parameter since the other ones follow from the detection of the crossings, has very different distributions. Noise produces a Gaussian-like distribution centered at $\mathcal{C}_\mathrm{SFMS}\sim8$ while the actual data has a bi-modal distribution with a larger peak at 0 crossings and a wider component centered around $\mathcal{C}_\mathrm{SFMS}\sim4$ instead. The reported preference for even values of crossings is not observed for noise-like fluctuations.

In Fig. \ref{fig:app-cross} we reproduce Fig. \ref{fig:cross} from the main text but instead of the ALL and RET galaxy groups we only show the results for MS and the same galaxies but with noise as the scatter of the SFMS. The results are similar to those of Fig. \ref{fig:app-dist}, with the differences between the actual results and noise are very apparent as the dependence on M$_\star$ either has large offsets or differences in the shape. The number of crossings remains virtually constant except for edge cases, while the deviation is mainly modulated by the scatter at that particular M$_\star$. 

 \subsection{How noise affects the fraction of time above the SFMS}
\begin{figure}[ht]%
\centering
\includegraphics[width=\linewidth]{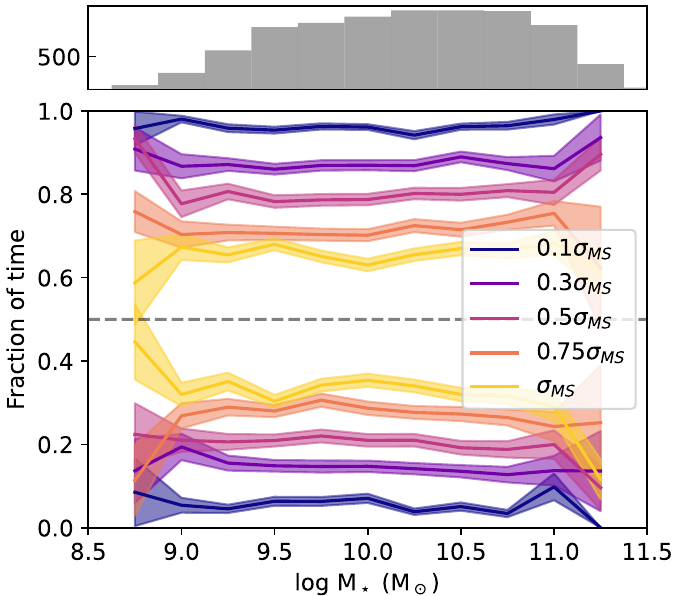}
\caption{Fraction of time spent above the SFMS for currently star-forming galaxies, having imposed that the current separation from the relation is maintained for all age values before adding noise. The colors correspond to the width of the random fluctuations that are added to this separation, as a factor of the scatter of the SFMS.} \label{fig:app-perc2}
\end{figure}

In Sec. \ref{sec:time} we showed that at least part of the scatter for the SFMS is due to the galaxies preferentially spending their time on one side of the relation. Galaxies appear to spend $\gtrsim$60\% of the time on the same side of the SFMS but this value could be significantly affected by noise, lowering it.
As should be expected, using random fluctuations the percentage of time spent on each side is almost perfectly 50\% for each mass bin, with the median value of the fraction for the M$_\star$ bins of 0.488 and a standard deviation of 0.05, which drops to 0.007 if we consider only the $10^{9-11}$ M$_\odot$ range, removing the least populated bins.
This reinforces the validity of the offset seen in Fig. \ref{fig:perc} as a physical result, showing that it cannot originate in noise or random uncertainties from the method.

Given that noise tends to produce even amounts of time above and below the SFMS, we can also assess how much lower the fraction of time spent above it we measure will be depending on the noise. To this end we perform a different inference regarding the residuals, assuming that the scatter of the SFMS is 100\% due to an offset in SFR values over time, with no crossings. As such, we take the separation from the SFMS that the galaxies currently have and impose the same separation for all age values in the past Gyr, thus ensuring fractions of time above the relation of 1 and 0 for galaxies currently above and below, respectively. Then we add noise to see how much this value is reduced depending on the noise level.

In Fig. \ref{fig:app-perc2} we show the result of adding Gaussian noise of different amplitudes to flat residuals where the current separation from the SFMS is preserved over time. We use five values of the amplitude scaled to the scatter of the relation at each M$_\star$, finding that higher noise amplitude makes the fraction of time spent above the relation closer to 0.5. A noise amplitude equal to the scatter reduces the fraction for galaxies currently above from 1 to $\gtrsim$0.6, similar to the values we observe. Of course, it is impossible to have both the separation and the noise simultaneously have the same value of the measured scatter as that would widen the scatter to $\sqrt{2}\sigma_{SFMS}$. Still, these results show that our measured $\gtrsim$60\% value for the fraction of time spent on one side of the SFMS is a lower limit. Our scatter measurements of 0.3-0.5 are about 0.1 dex wider than the range reported by the studies with the lowest measured scatters in the Local Universe \citep[e.g.,][]{Guo2013,Popesso2019}, though they are compatible with other determinations \citep[e.g.,][]{Brinchmann2004,Ilbert2015,Chang2015}. As such, we expect a possible underestimation of the fraction of time galaxies spend on one side but the offset should not be very large.

\end{appendix}

\end{document}